\documentclass[pra,a4paper,twocolumn,showpacs,superscriptaddress]{revtex4}
\usepackage{amsmath}
\usepackage{amsfonts}
\usepackage{graphicx}
\usepackage{longtable}
\newcommand{\be}{\begin{equation}}
\newcommand{\ee}{\end{equation}}

\newcommand{\ba}[1]{\left(\begin{array}{#1}}
\newcommand{\ea}{\end{array}\right)}
\begin{document}

\title{Purification and redistribution of entanglement via single local filtering} 
\author{K.O. Yashodamma }
\affiliation{Department of Physics, Kuvempu University, 
Shankaraghatta, Shimoga-577 451, India}
\author{P.J. Geetha }
\affiliation{Department of Physics, Kuvempu University, 
Shankaraghatta, Shimoga-577 451, India}
\author{Sudha}
\email{arss@rediffmail.com}
\affiliation{Department of Physics, Kuvempu University, 
Shankaraghatta, Shimoga-577 451, India}
\affiliation{Inspire Institute Inc., Alexandria, Virginia, 22303, USA.}
\date{\today}
\begin{abstract}
The effect of filtering operation with respect to purification and concentration of entanglement in quantum states are discussed in this paper. 
It is shown, through examples, that the local action of the filtering operator on a part of the composite quantum state allows for purification of the remaining part of the state. The redistribution of entanglement in the subsystems of a noise affected state is shown to be due to the action of local filtering on the non-decohering part of the system. The varying effects of the filtering parameter, on the entanglement transfer between the subsystems, depending on the choice of the initial quantum state is illustrated.  
\end{abstract}
\maketitle
\section{Introduction}
The concept of quantum entanglement\cite{n1,n2} is very useful in many areas like quantum teleportation\cite{n3,n4}, quantum cryptography\cite{n5}, quantum dense coding\cite{n6,n7} and quantum computation\cite{n8} but its fragility to environmental interaction is a cause of great concern for all its technological applications. It is well known that when a pure entangled state interacts with the environment, in addition to reduction in its entanglement, it loses its coherence also thus becoming a mixed state\cite{n8a}. Purification\cite{n9a} and concentration of entanglement\cite{n9b,n9c} have thus been issues of great importance. Several protocols including collective measurements, Local Quantum Operation with Classical Communication (LQCC) and local filtering operations are proposed to perform this task$[10--22]$. 

While it has been shown in Refs.~\cite{n9a,n9b,n9c,n9} that, the collective LOCC operations on an ensemble of bipartite states can increase the purity and entanglement of a smaller number of states by sacrificing the entanglement of the remaining states in the ensemble, the issue of entanglement purification in a single copy of a mixed bipartite state has been discussed in Refs.~\cite{n10,n11,n12}. The behaviour of local filtering operations in causing Bell violation and conversion of a mixed state to a Bell diagonal state are also examined\cite{n13a,n14,n13}.   
The importance of an optical entanglement filter and its applicability for entanglement manipulation is discussed in Refs.~\cite{n17,n18}. In addition to the experimental demonstration of the distillation of maximally entangled states from non maximally entangled inputs using partial polarizers\cite{n19}, optimal entanglement distillation from $2$-qubit mixed states under local filtering operations is also experimentally demonstrated\cite{exp}. The effect of local filtering operation on a part of the multiqubit state while the remaining part is subjected to noise is examined by M. Siomau and A.A. Kamli\cite{n16}. 
They have shown that entanglement can be probabilistically retrieved after Entanglement Sudden Death (ESD)\cite{esd,esd2}  using a local filter on  a single (atleast one) non-decohering qubit of a multiqubit state\cite{n16}. Continuing on this work\cite{n16}, we have examined here the behaviour of a local filter in causing purification of a part of the state, in addition to redistributing the entanglement in the subsystems. The filtering parameter dependence on both redistribution and purification is explicitly demonstrated for $3$-qubit pure states belonging to two SLOCC inequivalent families. Probabilistic retrieval of entanglement after ESD and the need for selecting a proper filter in order to have a maximum retrieval, depending on the state under consideration, is illustrated.     
   
The article is divided into three sections after a brief introduction in Section 1.  Section 2 gives an analysis of the effect of single local filtering on $3$-qubit pure states belonging to two different SLOCC classes. In Section 3, we have subjected a part of these states to a specific noise and examined the retrieval of entanglement due to local filtering on the other part of the system. The redistribution of entanglement between the subsystems resulting in an increase in the entanglement of one part owing to the reduction in the other part, is illustrated in both sections 2 and 3. Section 4 gives a concise summary of the results. 
\section{Single local filtering on 3-qubit pure states.} 
Filtering is a non-trace-preserving map  
that is seen to be capable of increasing entanglement with some probability\cite{n13a}. The filtering operation, in fact, represents a dichroic environment the extreme examples of which are polarizers\cite{n13a}.  While the local filtering operation can be realized using a Mach-Zehnder interferrometer\cite{mzi}, more efficient and feasible experimental realizations of a local filter are discussed in Refs. \cite{n19,exp}. The filtering operation can also be realized, using a linear optical set up, as a null-result weak measurement\cite{n20}.

 The filtering operation is represented in the computational basis as,  
\be
\label{f}
F = \sqrt{(1-k)} \vert 0 \rangle \langle 0 \vert + \sqrt{k} \vert 1 \rangle \langle 1 \vert 
\ee
Here $k$ is the filtering parameter and $0\leq k \leq 1$.  
 
To start with, we consider the $3$-qubit $W$ state  
\be
\label{ye}
\vert W \rangle_3 =\frac{1}{\sqrt{3}} \left[\vert 0 0 1\rangle +\vert 0 1 0\rangle +\vert 1 0 0 \rangle \right ].
\ee
and the application of filtering on only the first qubit  of the state results in
\be
\label{yc}
\vert W' \rangle_3 = \left[F \otimes I_2 \otimes I_2 \right]\vert W \rangle_3,
\ee
$I_2$ being the $2 \otimes 2$ identity matrix. Due to the non-trace preserving nature of the filtering operation, the physical state under consideration after filtering is given by 
\be
\label{tracew}
\rho'_W=\frac{\vert W' \rangle_3 \langle W'\vert}{\mbox{Tr} \left[ \vert W' \rangle_3 \langle W'\vert\right]}
\ee

It is well known that the entanglement between any two qubits of a $3$-qubit state can be quantified by concurrence\cite{n21,n22}. It is  given by $C= \mbox{max}\{0, \lambda_1-\lambda_2-\lambda_3-\lambda_4\}$, where $\lambda_i$ are square roots of the eigenvalues of the non-Hermitian matrix 
$\rho (\sigma_{y} \otimes \sigma_{y}) \rho^* (\sigma_{y} \otimes \sigma_{y})$ taken in decreasing order. 
As the $W$ state is symmetric under interchange of qubits, the initial concurrence of all its subsystems are equal, i.e., $C_{12}=C_{13}=C_{23}=\frac{2}{3}$.  After the first qubit is subjected to filtering, a redistribution of bipartite entanglement between the subsystems occurs. That is, the entanglement in the subsystem  $\rho'_{23}=\mbox{Tr}_1 \rho'_W$ increases with the  decrease in the entanglement of subsystems $\rho'_{12}=\mbox{Tr}_3 \rho'_W$, $\rho'_{13}=\mbox{Tr}_2 \rho'_W$ and vice versa. In fact, we have, 
\begin{eqnarray}
\label{red3}
\rho'_{12}=\rho'_{13}&=& \frac{1}{2-k}\ba{cccc} (1-k) & 0 & 0 & 0 \\ 0 & (1-k) & \sqrt{k(1-k)} & 0 \\ 0 & \sqrt{k(1-k)} & k & 0 \\ 0 & 0 & 0 & 0 \ea \nonumber \\ 
\rho'_{23}&=&\frac{1}{2-k} \ba{cccc} k & 0 & 0 & 0 \\ 0 &  (1-k)  &  (1-k)  & 0 \\ 0 &  (1-k)  &  (1-k) & 0 \\ 0 & 0 & 0 & 0 \ea 
\end{eqnarray} 
and 
\begin{eqnarray}
C'_{12}=C'_{13}&=&\mbox{Max} \left[0,\, 2\frac{\sqrt{k(1-k)}}{(2-k)} \right]; \nonumber \\
C'_{23}&=&\mbox{Max} \left[0,\, \frac{2(1-k)}{(2-k)} \right]. 
\end{eqnarray}
The redistribution of entanglement in the subsystems of $\vert W\rangle_3$, on its first qubit being subjected to filtering, is illustrated in Fig. 1. 
\begin{figure}[ht]
\centerline{\includegraphics* [width=2.4in,keepaspectratio]{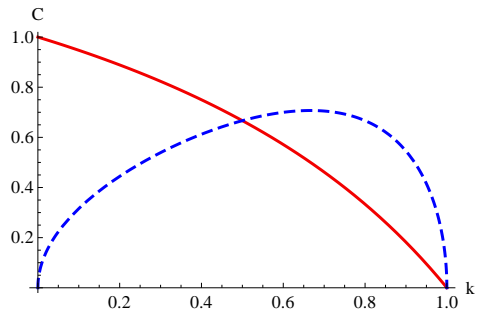}} 
\caption{The variation of concurrence of the subsystems $\rho'_{12}=\rho'_{13}$ (dashed line) and $\rho'_{23}$ (solid line) of the state $\rho'_W$ with the filtering parameter $k$;} 
\end{figure} 

Before examining the effect of the filtering operator as regards purification, we first define the  purity of a quantum state\cite{n8}. It is a scalar quantity defined as $\gamma = \mbox{Tr}(\rho^2)$.  For a pure state $\rho=\vert \psi \rangle \langle \psi \vert$ implies $\rho^2=\rho$ and hence $\mbox{Tr}(\rho^2)=\mbox{Tr}(\rho)=1$ for pure states. As $\mbox{Tr}(\rho^2)<1$ for mixed states, the quantity $1-\mbox{Tr}(\rho^2)$ quantifies the amount of mixedness in the state\cite{n8}. 

As the subsystems of the $W$ state (a symmetric entangled state) correspond to mixed states, we have  $\gamma_{12}=\gamma_{13}=\gamma_{23}=\frac{1}{3}$. The local operation of filtering on the first qubit destroys the symmetry of the state partially so that $\rho'_{12}=\rho'_{13}\neq \rho'_{23}$ (See Eq.(\ref{red3})). Also the purity of each subsystem becomes a function of the filtering parameter $k$. We have, 
\be
\label{purityw3}
\gamma'_{12}=\gamma'_{13}=\frac{2-(2-k)k}{(2-k)^2}; \ \  \gamma'_{23}=\frac{4-k(8-5k)}{(2-k)^2}
\ee
It can be seen that when $k=0$, the subsystem $\rho_{23}$ becomes
the maximally entangled Bell state $\frac{1}{\sqrt{2}} (\vert 01 \rangle + \vert 10 \rangle)$ and when $k=1$, it becomes the separable pure state $\vert 00 \rangle $ (See Eq. (\ref{red3})).  On the other hand, the subsystem $\rho_{13}$ becomes the separable pure state $\vert 10 \rangle $ at $k=1$. The variation of purity of the subsystems 
$\rho'_{12}$, $\rho'_{23}$ with the filtering parameter $k$ is as shown in Fig. 2. 
\begin{figure}[ht]
\centerline{\includegraphics* [width=2.4in,keepaspectratio]{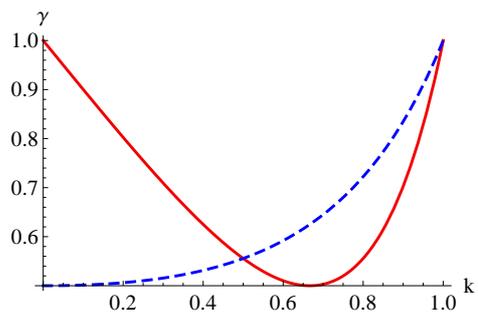}} 
\caption{The purity $ \gamma$ of $\rho'_{12}$ (dashed line) and $\rho'_{23}$ (solid line) of the state $\rho'_W$ as a function of the filtering parameter $k$. }
\end{figure} 
It can be readily seen that an increase in the entanglement of $\rho_{12}$ results in the decrease of $C'_{23}$, the concurrence of the subsystem $\rho'_{23}$, after filtering. 

The GHZ state\cite{ghz} is another symmetric multiqubit state of great importance possessing genuine multiparty entanglement and 
we carry out our analysis of the effect of single local filtering on a $3$-qubit GHZ state. The utmost fragility of GHZ state to removal of a qubit  
results in disentangled subsystems even after single local filtering, thus implying no essential redistribution of entanglement between them\cite{n16}. Still, the filtering operation is effective in increasing the purity of the initially mixed subsystems with $\gamma=\frac{1}{2}$. In fact, when $k=0$ and $k=1$, the local filtering on any one of the qubits completely destroys the entanglement in the GHZ state resulting in {\emph {pure}} (separable) subsystems with $\gamma=1$.

Another symmetric multiqubit state of interest is the equal superposition of $W$ and obverse $W$ states. We have the corresponding $3$-qubit state as, \begin{eqnarray*}
& & \vert W \bar{W}\rangle_3=\frac{1}{\sqrt{2}} \left[ \vert W \rangle_3 +\vert {\bar {W}} \rangle_3\right]   
\end{eqnarray*}
where $\vert {\bar {W}} \rangle_3=\frac{1}{\sqrt{3}}\left[\vert 1 1 0\rangle +\vert 1 0 1 \rangle +\vert 0 1 1 \rangle \right]$.
Both $3$-qubit GHZ state and $\vert W \bar{W}\rangle_3$ belong to the SLOCC family  ${\cal D}_{1,\,1,\,1}$ of symmetric states with three distinct Majorana spinors while $\vert W \rangle_3$ being a Dicke state, belongs to the family ${\cal D}_{2\, ,1}$, the class of three qubit symmetric states with two distinct spinors\cite{slocc}. Inspite of belonging to the same SLOCC family, the essential entanglement features of GHZ state and $\vert W \bar{W}\rangle$ differ in several aspects\cite{slocc}. In particular, $\vert W \bar{W}\rangle$ is not as much fragile to removal of qubits as the $3$-qubit GHZ state is and the subsystems remain entangled with concurrences $C_{12}=C_{13}=C_{23}=\frac{1}{3}$ even after the removal of a qubit.

On subjecting $\vert W \bar{W}\rangle_3$ to filtering on its first qubit, the symmetry among the subsystems is partially lost as in the case of $\vert W \rangle_3$ resulting in  $\rho'_{12}=\rho'_{13}\neq \rho'_{23}$. The subsystems $\rho'_{12}$(=$\rho'_{13}$) belonging to that part of the system containing the filter-affected qubit `$1$'  are seen to sacrifice their entanglement to increase the entanglement of $\rho_{23}$, the subsystem not directly affected by filtering. This behaviour can be readily seen through the variation of concurrences $C'_{12}$(=$C'_{13}$), $C'_{23}$ as a function of the filtering parameter $k$ as shown in  Fig. 3.  
\begin{figure}[ht]
\centerline{\includegraphics* [width=2.4in,keepaspectratio]{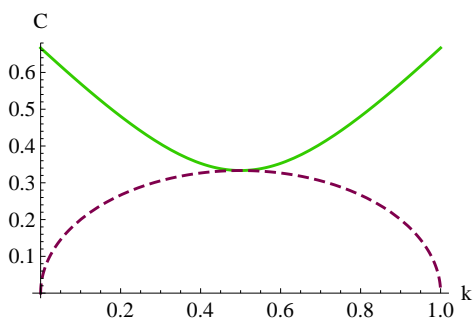}} 
\caption{The redistribution of entanglement between the subsystems $\rho'_{12}$, $\rho'_{13}$ and $\rho'_{23}$ of the filter-affected $\vert W \bar{W}\rangle_3$ state. Here $C'_{23}$ (solid line) is seen to increase over its initial value $C_{23}$ while $C'_{12}=C'_{13}\leq C_{12}=C_{13}$, ($C'_{12}, C'_{13}\rightarrow$\ dashed line)   over the whole range $0\leq k \leq 1$. }  
\end{figure}  

The subsystems of $\vert W \bar{W}\rangle_3$ being initially mixed, it is of interest to examine their purities after the local filtering action. We have, 
\be
\label{purwwbar}
\gamma'_{12}=\frac{1}{9} \left[7-2k(1-k)\right]=\gamma'_{13};\ \gamma'_{23}=\frac{1}{9} \left[9-10k(1-k)\right]
\ee  
It is readily seen that the subsystem $\rho'_{23}$ corresponds to a pure state when $k=0$ and $k=1$. In fact, they are entangled pure states with $C'_{23}=\frac{2}{3}$ both when $k=0$, $k=1$ (See Fig. 3).  The nature of variation of purity of the subsystems of the filter-affected $\vert W \bar{W}\rangle_3$ state is as shown in Fig. 4. 
\begin{figure}[ht]
\centerline{\includegraphics* [width=2.4in,keepaspectratio]{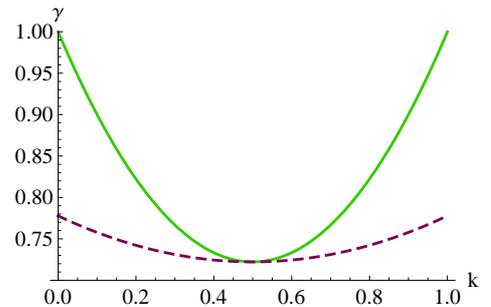}} 
\caption{The variation of purity $\gamma'_{12}$ (dashed line) and $\gamma'_{23}$ (solid line) of the subsystems belonging to the filter-affected $\vert W \bar{W}\rangle_3$ state as a function of the filtering parameter $k$.}
\end{figure}
Notice that when $k=\frac{1}{2}$, the filtering operator corresponds to identity operator thus implying $\gamma'_{12}=\gamma'_{13}=\gamma'_{23}\approx 0.7$ which is the initial purity of the subsystems. The entanglement in all the subsystems are also equal to their initial values when $k=0.5$ as can be seen through Figs. 1 and 3.

It is to be noted here that there is a distinct difference in the response of $\vert W \rangle_3$ and $\vert W \bar{W}\rangle_3$ to local filtering action. It is readily seen through Fig. 1 that the reduction in the entanglement of subsystems $\rho_{12}$, $\rho_{13}$ to contribute to  the increase in entanglement of subsystem $\rho_{23}$ of $\vert W \rangle_3$ is dependent on the filtering parameter $k$. While this contribution is largest when $k=0$, it goes on decreasing with the increase of $k$ and ceases when $k=0.5$. When $k>0.5$, a reverse trend occurs with the subsystem $\rho_{23}$ contributing to the increase in entanglement of $\rho_{12}$(=$\rho_{13}$) in a small range of the parameter $k$ (See Fig.1). But in the case of  $\vert W \bar{W}\rangle_3$, only the subsystems $\rho_{12}$, $\rho_{13}$ sacrifice their entanglement, to varying degrees, to the increase in entanglement of the subsystem $\rho_{23}$ and not vice versa. It is of interest to observe that the entanglement transfer from one subsystem (containing the `filtered' qubit 1) to the other (not directly affected by filtering) shows a symmetric behaviour about the value $k=0.5$ of the filtering parameter in the case of  $\vert W \bar{W}\rangle_3$ (See Fig. 3). Also in $\vert W \rangle_3$ the maximum entanglement concentration and purity in $\rho_{23}$ happens at $k=0$ whereas such a thing happens both at $k=0$ and $k=1$ for $\vert W \bar{W}\rangle_3$. Thus we can conclude that while a particular filter produces maximum entanglement concentration, purity in one state, it may not do so for the other state. The knowledge of the  filtering parameter\footnote{Though we have considered the filtering operation (See Eq. (\ref{f})) in the standard basis, one can choose a suitable measurement basis. The redistribution of entanglement, purity among the subsystems retain their nature provided all the measurements are done in the basis chosen for the filtering operation. The choice of a filtering parameter for maximum entanglement concentration, purity also depends on the choice of the measurement basis.} dependent nature of the redistribution of subsystem entanglement will help in choosing the needed `filter' for the chosen state, based on the particular task under consideration.   

\section{The effect of single local filtering on noise affected states.} 
In this section, we extend our analysis to states partially exposed to noisy environment. In fact, the retrieval of entanglement in the subsystems of multiqubit states, a part of which is subjected to noise and at least one qubit of the remaining part subjected to local filtering, is illustrated in Ref.~\cite{n16}. The Generalized Amplitude Damping (GAD) channel is used in Ref.~\cite{n16} as the chosen noise model and redistribution of entanglement, induced due to local filtering on a qubit, is shown to occur in the subsystems of noise affected $3$-qubit $W$ state, $4$-qubit cluster states. In view of the observation that depolarizing noise is more effective in causing sudden death of entanglement in quantum states\cite{rp,depygs}, we wish to examine the extent of entanglement retrieval, through single local filtering, in both $\vert W \rangle_3$ and $\vert W \bar{W}\rangle_3$ states subjected to depolarizing noise. 

A quantum noise that converts a qubit into a completely mixed
state with probability $p$ and leaves it untouched with probability
$1-p$ is the depolarizing noise\cite{n8}.
The Kraus operators corresponding to depolarizing noise are given in Ref.~\cite{n8} 
\begin{eqnarray}
k_1&=&\sqrt{1-p}
\ba{cc}
1 & 0 \\ 0 & 1
\ea;\ \  k_2= \sqrt{\frac{p}{3}}
\ba{cc}
0 & 1 \\ 1 & 0
\ea; \nonumber \\
k_3&=& \sqrt{\frac{p}{3}}
\ba{cc}
0 & i \\ -i & 0
\ea;\ k_4=\sqrt{\frac{p}{3}}
\ba{cc}
1 & 0 \\ 0 & -1
\ea.  
\end{eqnarray}
Here $p=1-e^{-\frac{\Gamma t}{2}}$ with $\Gamma$ being the decay constant of the depolarizing noise. 

The action of depolarizing noise on any two qubits (say `$2$'and  $3$')  of a $3$-qubit state $\rho$  is represented by 
\be
\rho'= \sum_{i,j}  s_{ij} \rho s_{ij}^{\dagger}
\ee
with   
\[
s_{ij} = I_2 \otimes k_{i} \otimes k_{j}; \ \ \sum_{i,j}{s_{ij}^{\dagger} s_{ij}} = I_8
\]
$I_2$, $I_8$ being the $2\times 2$ and $8\times 8$ identity matrices respectively. In addition to delaying the sudden death of entanglement caused in the subsystem $\rho'_{23}$,  an increase in entanglement can also be seen to occur on subjecting $\rho'$ to local filtering on its first qubit. This action is represented by 
\be
\rho''= \frac{\left( F\otimes I_2 \otimes I_2  \right)  \rho'  \left( F\otimes I_2 \otimes I_2  \right)^\dagger}{\mbox{Tr}\,\left( F\otimes I_2 \otimes I_2  \right)  \rho'  \left( F\otimes I_2 \otimes I_2  \right)^\dagger} 
\ee
The combined action of depolarizing noise and local filtering on $\vert W \rangle_3$ and $\vert W \bar{W}\rangle_3$ is seen to retain the partial symmetry in both the states resulting in $\rho''_{12}=\rho''_{13}\neq \rho''_{23}$. The variation with respect to time, of the concurrence of the subsystem  
$\rho''_{23}$  (belonging to the noise affected state $\vert W \rangle_3$) for different values of the filtering parameter $k$ is shown in Fig. 5.  
The corresponding graph for the state $\vert W \bar{W}\rangle_3$ is given in Fig. 6.  
\begin{figure}[ht] 
\centerline{\includegraphics* [width=2.4in,keepaspectratio]{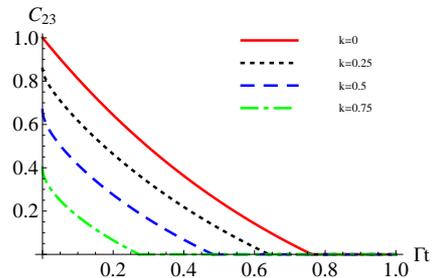}} 
\caption{The variation of concurrence of the subsystem  $\rho''_{23}$, of the noise affected state $\vert W \rangle_3$  after local filtering on its 1st qubit, with the dimensionless time parameter $\Gamma t$ for different values of the filtering parameter $k$.} 
\end{figure}   
The delay in the onset of ESD as well as increase in the entanglement in $\rho_{23}$ for $k=0$, compared to its initial value (when k=0.5), is readily seen in Fig 5. Such a behaviour is exhibited for $0\leq k <0.5$. In fact, when $0.5<k\leq 1$, there is a reduction in entanglement of $\rho''_{23}$ as well as early onset of ESD (See Fig. 5).
\begin{figure}[ht]
\centerline{\includegraphics* [width=2.4in,keepaspectratio]{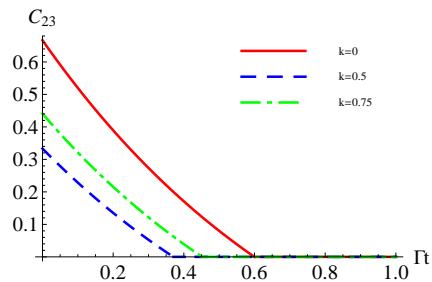}} 
\caption{Temporal variation of concurrence $C''_{23}$ of the subsystem $\rho''_{23}=\mbox{Tr}_1 \rho''_{W\bar{W}}$ for different values of the filtering parameter $k$. For all values of $k$, gain in the entanglement of $\rho_{23}$ and delay in the onset of ESD (compared to that without filtering) is observed.}. 
\end{figure}  
As both the qubits of the subsystem $\rho_{23}$ are subjected to noise, it is not difficult to see that the entanglement of this subsystem vanishes earlier than for the subsystems $\rho_{12}$(=$\rho_{13}$).  Figs. 7 and 8 depicting the temporal variation of $C''_{12}$(=$C''_{13}$), for different values of the filtering parameter $k$, indicate this feature.
\begin{figure}[ht]
\centerline{\includegraphics* [width=2.4in,keepaspectratio]{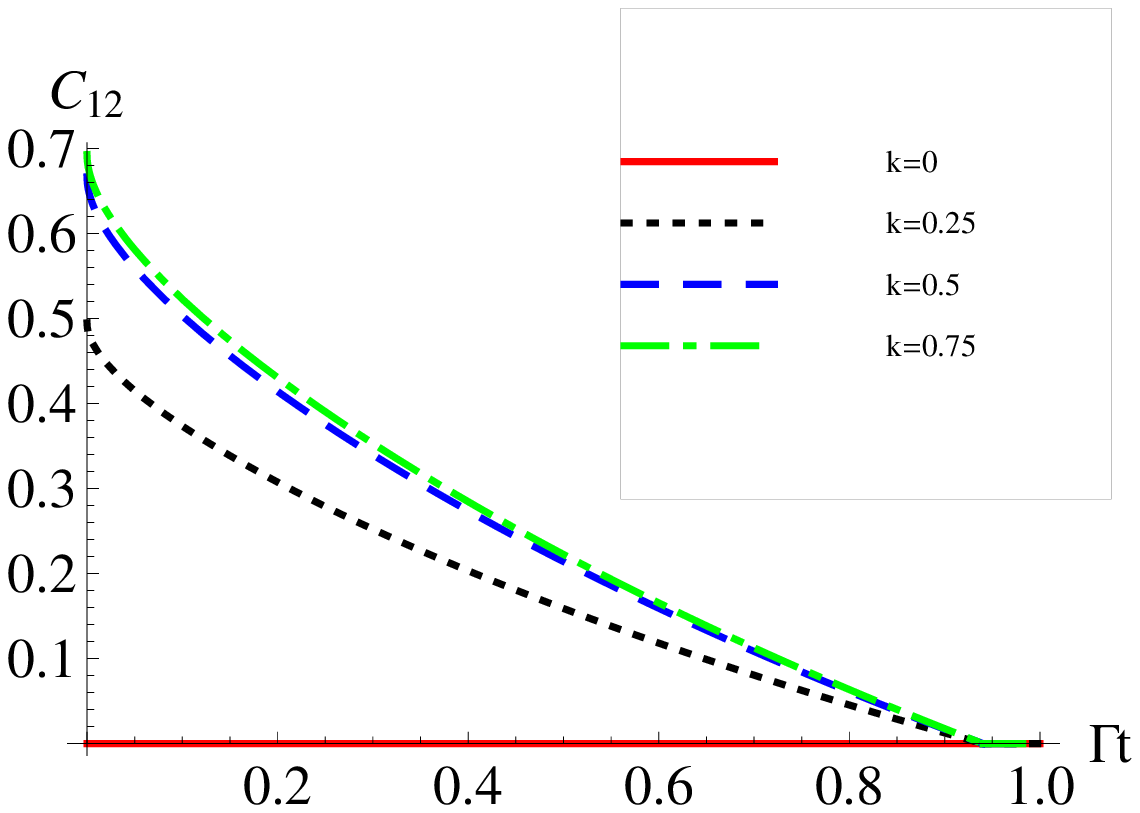}} 
\caption{Variation with respect to time of the  concurrence of $\rho''_{12}=\mbox{Tr}_3 \rho''_{W}$ for different values of the filtering parameter $k$.}
\end{figure}
\begin{figure}[ht]
\centerline{\includegraphics* [width=2.4in,keepaspectratio]{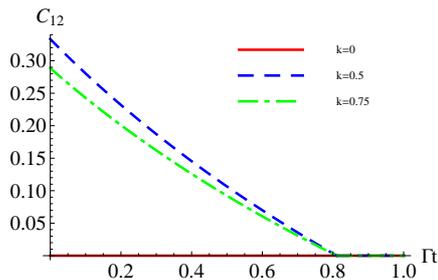}} 
\caption{Variation with respect to time of the  concurrence of the subsystem $\rho'_{12}=\mbox{Tr}_3 \rho''_{W{\bar W}}$ for different values of the filtering parameter $k$}.  
\end{figure} 
The filtering parameter dependent redistribution of entanglement between the subsystems $\rho''_{12}$ and $\rho''_{23}$ of $\vert W\rangle_3$ is evident through a closer look at Figs. 5 and 7. In particular, the largest transfer of entanglement happens when $k=0$ with $C''_{12}=0$ and $C''_{23}=1$ at time $t=0$. Similar and much simpler redistribution is seen through Figs. 6 and 8, for the case of $\vert W\bar{W}\rangle$. 

It is to be mentioned here that in all cases of entanglement transfer between the subsystems, the process is indeed {\emph{ probabilistic}}. 
The probability of `retrieval of entanglement' is, in fact, given by 
\be
\label{prob}
P=\mbox{Tr}\,\left( F\otimes I_2 \otimes I_2  \right)  \rho'  \left( F\otimes I_2 \otimes I_2  \right)^\dagger,  \ 0\leq P\leq 1 ;
\ee
That is, if we consider an ensemble of states $\rho$, $P$ represents the fraction of the states that undergo the effects of the filtering operation. It represents the efficiency of the filter $F$. 

While we have considered pure states for our analysis of the effects of filtering action, similar effects are expected to hold good even for mixed states. For a mixed state, due to its possible {\emph {apriori}} interaction with the surroundings (which makes the state lose its coherence) the nature of ESD depends on this unknown noise along with the applied noise.  Due to the action of this {\emph {apriori}} noise, ESD due to a particular applied noise may happen faster in entangled mixed states than in pure states. In order to delay the onset of ESD, one may have to choose a `suitable' filter depending on the mixed state under consideration. Also, it is to be noticed that in both the states that we have considered, maximum retrieval of entanglement happens with suitable choice of the filter. For instance, $k=0$ is suitable for $\vert W \rangle_3$ whereas 
both $k=0$, $k=1$ are suitable for $\vert W \bar{W}\rangle_3$ as we had observed earlier. Thus, we may conjecture that for every pure state, it is possible to find a suitable filter that causes maximum entanglement redistribution/retrieval after ESD. 

At this stage, one can readily conclude that local filtering action on atleast one of the qubits  is responsible for redistribution of entanglement among the subsystems of a $3$-qubit state. We assert that an extension of the results to multiqubit systems with $N>3$ is evident, though one has to consider all possible partitions of the state, for a complete analysis.  It is natural to expect that if more than one qubit is available for filtering action, the pairwise entanglement in the remaining non-decohering qubits will increase considerably\cite{n16}.   

Before concluding, we wish to mention here that this work is a sequel to the work of Ref. ~\cite{n16}. While the varying effects of a filter depending on the state under consideration and the role of the filtering parameter in defeating sudden death effectively has not been explicitly analyzed in Ref.~\cite{n16}, we have carried out such an analysis in this work. The `purifying' action of filtering operator on the subsystems of a multiqubit state is another additional feature that we have explicitly brought out here. Owing to the experimental implementations of a local filter\cite{n19,exp} in entanglement distillation protocols, we believe that retrieval of entanglement in quantum states after ESD will also be a reality in the near future. 

\section{Conclusion} 
We have shown here that it is possible to increase the subsystem entanglement of pure quantum states by single local filtering on a part of the system. The redistribution in the entanglement between the subsystems, induced due to local filtering operation, is shown to be the cause for gain in entanglement in the other part of the system. An increase in the purity of one subsystem owing to the decrease in the other is also shown to be caused due to local filtering.  Both purification and entanglement concentration in the subsystems of the entangled initial states are seen to be probabilistic in nature.  The dependence of the nature of entanglement transfer on the filtering parameter is clearly brought out and its consequences are discussed. These results are exemplified by examining $3$-qubit $W$ state, a superposition of $W$ and obverse $W$ states, both belonging to different SLOCC classes. The results are extendible to multiqubit states with $N>3$ and this work is hoped to shed more light on the  significance of local filtering action on quantum states. The technologically feasible experimental realizations of filtering operations\cite{n19,exp} also enrich the potentialities of this field.

\section*{Acknowledgments}

K.O. Yashodamma  and P.J. Geetha  acknowledge the support of Department of Science and Technology (DST), Govt. of India through the award of INSPIRE fellowship.

\vspace*{-5pt}   


\begin{thebibliography}{0}
\bibitem{n1} E. Schr\"{o}dinger, {\it Naturewissenschaften}, {\bf 23} (1935)  807.  
\bibitem{n2} A. Einstein, B. Podolsky, N. Rosen,  {\it Phys. Rev.} {\bf 47} (1935) 777.
\bibitem{n3} C. H. Bennett {\it et al}., {\it Phys. Rev. Lett.} {\bf 70} (1993) 1895.
\bibitem{n4} L. Vaidman, {\it Phys. Rev. A.} {\bf 49} (1994) 1473.
\bibitem{n5} A. K. Ekert, {\it Phys. Rev. Lett.} {\bf 67} (1991)  661.
\bibitem{n6} C. H. Bennett and S. J. Wiesner, {\it Phys. Rev. Lett.} {\bf 69} (1992) 2881.
\bibitem{n7} A. Harrow,  P. Hayden and D. Leung, {\it Phys. Rev. Lett.} {\bf 92} (2004) 187901.
\bibitem{n8} M. A. Nielsen, I. I. Chuang, {\it Quantum Computation and Quantum Information} (Cambridge University Press, Cambridge, 2000). 
\bibitem{n8a} H.-P. Breuer, F. Petruccione, {\it Theory of Open Quantum Systems} (Oxford University Press, Oxford, 2002) 
\bibitem{n9a} C. H. Bennett {\it et al}., {\it Phys. Rev. Lett.} {\bf 76} (1996) 722.
\bibitem{n9b} C. H. Bennett {\it et al}.,  {\it Phys. Rev. A.} {\bf 53} (1996) 2046. 
\bibitem{n9c} H. K. Lo  and  S. Popescu, quant-Ph/9707038v2.
\bibitem{n9}  M. Horodecki, P. Horodecki and R. Horodecki, {\it Phys. Rev. Lett.} {\bf 78} (1997) 574.
\bibitem{n10} N. Linden, S. Massar and  S. Popescu {\it Phys. Rev. Lett.} {\bf 81}, (1998) 3279    
\bibitem{n11}  A. Kent, {\it Phys. Rev. Lett.} {\bf 81} (1998) 2839.
\bibitem{n12} L.-X. Cen, F.-L. Li and S.Y. Zhu, {\it Phys. Lett. A} {\bf 275} (2000) 368.
\bibitem{n13a} N. Gisin, {\it Phys. Lett. A} {\bf 210} (1996) 151. 
\bibitem{n14} L-X Cen {\it et al}., {\it Phys. Rev. A} {\bf 65} (2002) 052318. 
\bibitem{n14a} J.-W. Pan {\it et al}., {\it Nature} (2001) 
{\bf 410}, 1067.
\bibitem{n14b} T. Yamamoto {\it et al}.,  {\it Nature} {\bf 421} (2003) 343.
\bibitem{n15} F. Verstraete, J. Dehaene  and  B. DeMoor, {\it Phys. Rev. A}  {\bf 64} (2006) 010101 .
\bibitem{n16} M. Siomau  and A. A. Kamli,  {\it Phys. Rev. A} {\bf 86} (2012) 032304 . 
\bibitem{n13} F. Verstraete and M. M. Wolf, {\it Phys. Rev. Lett.} {\bf 89} (2002) 170401.
\bibitem{n17} H. F. Hofmann   and S. Takeuchi, {\it Phys. Rev. Lett.}, {\bf 88} (2002) 147901.
\bibitem{n18} R. Okamoto {\it et.al}., {\it Science} {\bf 323} (2009) 483. 
\bibitem{n19} P. G. Kwiat {\it et.al}., {\it  Nature} {\bf 409} (2001)  1014.
\bibitem{exp} Z.-W. Wang {\it et al}., {\it Phys. Rev. Lett.}, {\bf 96} (2006) 220505.  
\bibitem{esd} L. Di\'{o}si,  {\it Lect. Notes. Phys.} {\bf 622} (2003) 157. 
\bibitem{esd2} T. Yu   and  J. H. Eberly,  {\it Phys. Rev. Lett.} {\bf 97} (2006) 140403. 
\bibitem{mzi} C. Zhang, {\it Quantum Inf. Comput.} {\bf 4}, (2004) 196.  
\bibitem{n20} Q. Sun {\it et.al}.,  {\it Phys. Rev. A.} {\bf 82} (2010) 032323.
\bibitem{n21} S. Hill   and  W. K. Wootters, {\it Phys. Rev. Lett.} {\bf 78} (1997) 5022. 
\bibitem{n22} W. K. Wootters, {\it Phys. Rev. Lett.} {\bf 80} (1998) 2245. 
\bibitem{ghz} W. \"{D}ur,  G. Vidal and  J. I. Cirac, {\it Phys. Rev. A} {\bf 62} (2000) 062314.
\bibitem{slocc} A. R. Usha Devi, Sudha and  A. K. Rajagopal,  {\it Quantum Inf Process} {\bf 11} (2012) 685. 
\bibitem{rp} K. O. Yashodamma and Sudha, {\it Results in Physics} {\bf 3} (2013) 41. 
\bibitem{depygs} K. O. Yashodamma, P. J. Geetha and Sudha, quant-Ph/1310.0715.
\end{thebibliography}
\end{document}